\documentclass[preprint,aps,showpacs]{revtex4}
\usepackage[dvips]{graphicx}

\begin{document}

\newcommand{\dfrac}[2]{\frac{\displaystyle #1}{\displaystyle #2}}
\preprint{VPI--IPPAP--02--11}

\title{The NuTeV Anomaly, Neutrino Mixing, and a Heavy Higgs Boson}
\author{Will~Loinaz\footnote{electronic address: loinaz@alumni.princeton.edu}}
\affiliation{Department of Physics, Amherst College, Amherst MA 01002}
\author{Naotoshi~Okamura\footnote{electronic address: nokamura@vt.edu},
Tatsu~Takeuchi\footnote{electronic address: takeuchi@vt.edu}}
\affiliation{Institute for Particle Physics and Astrophysics,
Physics Department, Virginia Tech, Blacksburg, VA 24061}
\author{L.~C.~R.~Wijewardhana\footnote{electronic address: rohana@physics.uc.edu}}
\affiliation{Department of Physics, University of Cincinnati, Cincinnati OH 45221-0011, and\\
the Institute of Fundamental Studies, Kandy, Sri Lanka}

\date{October 12, 2002; revised December 6, 2002; January 16, 2003}

\begin{abstract}
Recent results from the NuTeV experiment at Fermilab and 
the deviation of the $Z$ invisible width, measured at LEP/SLC,
from its Standard Model (SM) prediction suggest the suppression of
neutrino-$Z$ couplings.
Such suppressions occur naturally in models which mix the neutrinos
with heavy gauge singlet states.
We postulate a universal suppression of the $Z\nu\nu$ couplings by a 
factor of $(1-\varepsilon)$ and perform a fit to the $Z$-pole and NuTeV 
observables with $\varepsilon$ and the oblique correction parameters 
$S$ and $T$.  Compared to a fit with $S$ and $T$ only, inclusion of
$\varepsilon$ leads to a dramatic improvement in the quality of the fit.
The values of $S$ and $T$ preferred by the fit can be obtained within 
the SM by a simple increase in the Higgs boson mass.
However, if the $W$ mass is also included in the fit, a non-zero
$U$ parameter becomes necessary which cannot be supplied within the SM.
The preferred value of $\varepsilon$ suggests that the seesaw mechanism
may not be the reason why neutrinos are so light.
\end{abstract}


\maketitle

\section{Introduction}

The \textit{Neutrinos at the Tevatron} (NuTeV) experiment \cite{Zeller:2001hh} 
at Fermilab has measured the ratios of neutral to charged current events in 
muon (anti)neutrino -- nucleon scattering:
\begin{eqnarray}
R_\nu 
& = & \dfrac{ \sigma(\nu_\mu N \rightarrow \nu_\mu X) }
            { \sigma(\nu_\mu N \rightarrow \mu^-   X) }
\;=\; g_L^2 + r g_R^2\;, \cr
R_{\bar{\nu}}
& = & \dfrac{ \sigma(\bar{\nu}_\mu N \rightarrow \bar{\nu}_\mu X) }
            { \sigma(\bar{\nu}_\mu N \rightarrow \mu^+         X) }
\;=\; g_L^2 + \dfrac{g_R^2}{r}\;,
\end{eqnarray}
where
\begin{equation}
r = \dfrac{ \sigma( \bar{\nu}_\mu N \rightarrow \mu^+ X) }
          { \sigma( \nu_\mu       N \rightarrow \mu^- X) }
\sim \frac{1}{2}\;,
\end{equation}
and has determined the parameters $g_L^2$ and $g_R^2$ \cite{LlewellynSmith:ie}
to be
\begin{eqnarray}
g_L^2 & = & 0.30005 \pm 0.00137\;, \cr
g_R^2 & = & 0.03076 \pm 0.00110\;.
\label{nutev}
\end{eqnarray}
The Standard Model (SM) predictions of these parameters 
based on a global fit to non-NuTeV data, cited as 
$[g_L^2]_\mathrm{SM}=0.3042$ and 
$[g_R^2]_\mathrm{SM}=0.0301$ in Ref.~\cite{Zeller:2001hh},
differ from the NuTeV result by $3\sigma$ in $g_L^2$.
This disagreement between NuTeV and the 
SM (as determined by non-NuTeV data) is sometimes referred to
as the NuTeV `anomaly' \cite{Davidson:2001ji}.

Various suggestions have been forwarded as the cause of this 
discrepancy \cite{Babu:1991ny,DeGouvea:2001mz,
Giunti:2002nh,Barshay:2002hu,Ma:2001tb,Miller:2002xh,
Kovalenko:2002xe,Kumano:2002ra}.
These include theoretical uncertainties due to QCD effects, 
tree level and loop effects resulting from new physics, and
nuclear physics effects.
We refer the reader to Ref.~\cite{Davidson:2001ji}
for a comprehensive review.
In this paper, we investigate one explanation which is
particularly attractive in its simplicity and economy.

Note that the NuTeV value of $g_L^2$ is \textit{smaller} than the SM 
prediction.  This is because the ratios
$R_\nu$ and $R_{\bar{\nu}}$ were both smaller than expected,
\textit{i.e.} the neutral current events were not
as numerous as predicted by the SM when compared to the charged
current events. 
In addition, the invisible width of the $Z$ measured at 
the CERN/SLAC $e^+e^-$ colliders LEP/SLC is 
also known to be $2\sigma$ below
the SM prediction \cite{LEP/SLC:2002}.
Both observations suggest that the
couplings of the neutrinos to the $Z$ boson are suppressed 
with respect to the SM.

Suppression of the $Z\nu\nu$ couplings occurs naturally in
models which mix the neutrinos with heavy gauge singlet states 
\cite{Babu:1991ny,DeGouvea:2001mz,Gronau:ct,Bernabeu:gr,
Chang:1994hz,Marciano:1999ih}.
For instance, if the $SU(2)_L$ active neutrino $\nu_L$
is a linear combination of two mass eigenstates with
mixing angle $\theta$, 
\begin{equation}
\nu_L = (\cos\theta) \nu_\mathrm{light} + (\sin\theta) \nu_\mathrm{heavy}\;,
\end{equation}
then the $Z\nu\nu$ coupling will be suppressed by a factor of
$\cos^2\theta$ if the heavy state is too massive to be created
in the interaction.  Note that the $W\ell\nu$ coupling
will also be suppressed by a factor of $\cos\theta$.

In general, if the $Z\nu\nu$ coupling of a particular neutrino 
flavor is suppressed by a factor of $(1-\varepsilon)$, 
then the $W\ell\nu$ coupling of the same flavor will
be suppressed by a factor of $(1-\varepsilon/2)$.
For simplicity, and to preserve lepton universality, 
assume that the suppression parameter $\varepsilon$ is common to 
all three generations.
The theoretical values of $R_\nu$ and $R_{\bar{\nu}}$
are reduced by a factor of $(1-\varepsilon)$, since their numerators
are suppressed over their denominators, and the invisible width of 
the $Z$ is reduced by a factor of $(1-2\varepsilon)$.
At first glance, then, it seems that neutrino mixing may 
explain both the NuTeV and invisible width discrepancies.

However, more careful consideration appears to veto this possibility.
One of the inputs used in calculating the
SM predictions is the Fermi constant $G_F$,
which is extracted from the muon decay constant $G_\mu$.
Suppression of the $W\ell\nu$ couplings leads to the correction
\cite{Babu:1991ny,Gronau:ct,Bernabeu:gr,Chang:1994hz}
\begin{equation}
G_F = G_\mu (1+\varepsilon)\;,
\label{GFshift}
\end{equation}
which will affect \textit{all} SM predictions.
Ref.~\cite{Davidson:2001ji} concludes that 
a value of $\varepsilon$ large enough to explain the NuTeV anomaly
would affect the predictions of the 
SM (with a conventional light Higgs scalar) to an extent that the 
excellent agreement between the SM and $Z$-pole observables would be lost.

Notice, though, that the Fermi constant $G_F$ always appears
multiplied by the $\rho$-parameter in neutral current amplitudes.
Thus, the predictions for $Z$-pole observables will be undisturbed
if a shift in $G_F$ is compensated by a shift in $\rho$.
Such shifts can arise via oblique corrections 
due to new physics \cite{Peskin:1990zt}.

In the following, we perform a fit to the $Z$-pole and
NuTeV data with the oblique correction parameters $S$ and $T$
as well as the $Z\nu\nu$ suppression parameter $\varepsilon$.
Oblique corrections alone cannot explain
the NuTeV and invisible width discrepancies.  
However, fits involving the parameter $\varepsilon$ 
show excellent agreement with both the $Z$-pole and NuTeV data.

The question then, is: ``What physics would account for
the values of $S$ and $T$ that allow for a non-zero $\varepsilon$?''
A simple (though perhaps not unique) solution is the SM itself
with a heavy Higgs boson.
Indeed, we show that if the Higgs boson mass is allowed to be large, then
the $Z$-pole and NuTeV data can be fit by $\varepsilon$ alone.

The preferred value of $\varepsilon$ suggests large mixing
angles between the active neutrinos and the heavy sterile states.
This may rule out the seesaw mechanism \cite{seesaw} as the
explanation of the smallness of neutrino masses.

If the $W$ mass is included in the fit, we must also 
introduce the oblique correction parameter $U$. The preferred value 
of $U$ is virtually independent of the reference Higgs boson mass 
and cannot be supplied within the SM.
We discuss what type of new physics may generate
the required value of $U$.

\section{The Data}

\begin{table}
\begin{tabular}{c|ccccc}
\hline
& $M_Z$ 
& $\Gamma_Z$ 
& $\sigma_\mathrm{h}^0$ 
& $R_\ell^0$ 
& $A_\mathrm{FB}^{0,\ell}$ \\
\hline
$M_Z$ 
& $\phantom{-}1.000$ & & & & \\
$\Gamma_Z$ 
& $-0.023$ & $\phantom{-}1.000$ & & & \\
$\sigma_\mathrm{h}^0$ 
& $-0.045$ & $-0.297$ & $\phantom{-}1.000$ & & \\
$R_\ell^0$ 
& $\phantom{-}0.033$ & $\phantom{-}0.004$ 
& $\phantom{-}0.183$ & $\phantom{-}1.000$ & \\
$A_\mathrm{FB}^{0,\ell}$ 
& $\phantom{-}0.055$ & $\phantom{-}0.003$ 
& $\phantom{-}0.006$ & $-0.056$ & $\phantom{-}1.000$ \\
\hline
\end{tabular}
\caption{The correlation matrix of the $Z$--lineshape parameters from LEP.}
\label{ZlineshapeCorr}
\end{table}

We begin by listing the data which we will use in our analysis.

The $Z$ lineshape parameters from LEP/SLC assuming lepton
universality are (Ref.~\cite{LEP/SLC:2002}, page 8)
\begin{eqnarray}
M_Z                    & = & 91.1875 \pm 0.0021\,\mathrm{GeV}\;, \cr
\Gamma_Z               & = & 2.4952 \pm 0.0023\,\mathrm{GeV}\;, \cr
\sigma_\mathrm{h}^0    & = & 41.540 \pm 0.037\,\mathrm{nb}\;, \cr
R_\ell^0               & = & 20.767 \pm 0.025\;, \cr
A_\mathrm{FB}^{0,\ell} & = & 0.0171 \pm 0.0010\;,
\label{Zlineshape}
\end{eqnarray}
with the correlation matrix shown in Table~\ref{ZlineshapeCorr}.
Of these parameters, $M_Z$ is used as input to calculate the
SM predictions, and $R_\ell^0$ is used to fix the QCD coupling
constant $\alpha_s(M_Z)$.   The remaining three are equivalent to
(Ref.~\cite{LEP/SLC:2002}, pages 8, 9, and 146)
\begin{eqnarray}
\Gamma_\mathrm{lept}
& = & 83.984 \pm 0.086\,\mathrm{MeV}\;, \cr
\Gamma_\mathrm{inv}/\Gamma_\mathrm{lept}
& = & 5.942 \pm 0.016 \;, \cr
\sin^2\theta_{\mathrm{eff}}^\mathrm{lept}
& = & 0.23099 \pm 0.00053 \;.
\label{DATA2}
\end{eqnarray}
There is a correlation of $0.17$ between $\Gamma_\mathrm{lept}$ and
$\Gamma_\mathrm{inv}/\Gamma_\mathrm{lept}$ while other correlations
are negligible.
The SM prediction for the $Z$ invisible
width is (Ref.~\cite{LEP/SLC:2002}, page 9)
\begin{equation}
\Gamma_\mathrm{inv}/\Gamma_\mathrm{lept}
= 5.9736 \pm 0.0036\;.
\end{equation}
As mentioned in the introduction, this is $2\sigma$ above the 
experimental value.

The remaining observables from LEP and SLC, {\it i.e.} various asymmetries
and ratios of partial widths, can be interpreted as measurements of 
$\sin^2\theta_{\mathrm{eff}}^\mathrm{lept}$ in the absence of vertex
corrections from new physics \cite{Takeuchi:1994zh}. 
The average of the values obtained from $\tau$ polarization,
$A_\mathrm{LR}$, $A_\mathrm{FB}^{0,b}$, $A_\mathrm{FB}^{0,c}$, and
$\langle Q_\mathrm{FB}\rangle$, and 
that obtained from $A_\mathrm{FB}^{0,\ell}$ above, 
is (Ref.~\cite{LEP/SLC:2002}, page 146)
\begin{equation}
\sin^2\theta_{\mathrm{eff}}^\mathrm{lept}
= 0.23148 \pm 0.00017\;.
\label{DATA3}
\end{equation}
It should be noted that the agreement among the
values of $\sin^2\theta_{\mathrm{eff}}^\mathrm{lept}$ obtained from
the six observables is not good.  The $\chi^2/\mathrm{d.o.f.}$ 
associated with the average is $10.2/5$.  However, since neither
oblique corrections nor neutrino mixing has any effect on the
quality of the agreement, we will just use
the average value, Eq.~(\ref{DATA3}), as representative of these
measurements.  The observables not included in this 
average, such as $R_b$ and $R_c$,
depend only weakly on $\sin^2\theta_{\mathrm{eff}}^\mathrm{lept}$
and carry little statistical significance \cite{RbComment}.
Indeed, we have also performed an analysis with all the asymmetries and
heavy flavor data, together with their correlations, included separately
and have confirmed that doing so does not change the conclusions of
this paper.

For $g_L^2$ and $g_R^2$, we take the values
in the 1998 Particle Data Book \cite{PDB:1998}, which are based on
all experiments prior to NuTeV,
\begin{eqnarray}
g_L^2 & = & 0.3009 \pm 0.0028\;, \cr
g_R^2 & = & 0.0328 \pm 0.0030\;,
\end{eqnarray}
and average them with the NuTeV values, Eq.~(\ref{nutev}).
We obtain,
\begin{eqnarray}
g_L^2 & = & 0.3002 \pm 0.0012\;, \cr
g_R^2 & = & 0.0310 \pm 0.0010\;.
\label{DATA1}
\end{eqnarray}
The correlation is negligible.

We will also use the $W$ mass later in the analysis.
The world average from $p\bar{p}$ experiments and LEP2 is
(Ref.~\cite{LEP/SLC:2002}, page 150)
\begin{equation}
M_W = 80.449 \pm 0.034\,\mathrm{GeV}\;.
\label{Wmass}
\end{equation}

The SM predictions are calculated with ZFITTER v6.36 \cite{ZFITTER:2001}
and the formulae in Ref.~\cite{Marciano:pb}.
The inputs to ZFITTER are the $Z$ mass from Ref.~\cite{LEP/SLC:2002},
listed in Eq.~(\ref{Zlineshape}), the top mass from Ref.~\cite{CDF/D0:2000},
\begin{equation}
M_\mathrm{top} = 174.3 \pm 5.1\,\mathrm{GeV}\;,
\end{equation}
and the hadronic contribution to the running of the QED coupling constant
from Ref.~\cite{Burkhardt:2001xp},
\begin{equation}
\Delta\alpha_\mathrm{had}^{(5)}(M_Z) = 0.02761 \pm 0.00036\;.
\end{equation}
The QCD coupling constant is chosen to be
\begin{equation}
\alpha_s(M_Z) = 0.119\;.
\end{equation}
All our observables are either purely leptonic, hence QCD corrections
only enter at the two loop level, or ratios of cross sections in which the
leading order QCD corrections cancel.
Consequently, the results of the fits depend only very weakly on this choice.
Finally, the Higgs mass is varied from 115~GeV, the lower bound from
direct searches \cite{PDB:2002}, up to 1000~GeV.

\begin{table}
\begin{tabular}{ccc}
Observable & SM prediction & Measured Value \\
\hline
$\Gamma_\mathrm{lept}$ & 
$83.998$~MeV & 
$83.984\pm 0.086$~MeV \\
$\Gamma_\mathrm{inv}/\Gamma_\mathrm{lept}$ &
$5.973$ &
$5.942 \pm 0.016$ \\
$\sin^2\theta_\mathrm{eff}^\mathrm{lept}$ & 
$0.23147$ &
$0.23148\pm 0.00017$ \\
$g_L^2$ & 
$0.3037$ &
$0.3002 \pm 0.0012$ \\
$g_R^2$ &
$0.0304$ &
$0.0310 \pm 0.0010$ \\
$M_W$ &
$80.375$ &
$80.449\pm 0.034$~GeV \\
\hline
\end{tabular}
\caption{The observables used in this analysis.
The SM predictions are for inputs of
$M_\mathrm{top}=174.3\,\mathrm{GeV}$,
$M_\mathrm{Higgs}=115\,\mathrm{GeV}$,
$\alpha_s(M_Z) = 0.119$, and
$\Delta\alpha_\mathrm{had}^{(5)} = 0.02761$.
They differ from the SM predictions cited in the text since the
global fit prefers a slightly different set of input values.}
\end{table}

\section{The Corrections}

We now consider the effect of the oblique correction parameters 
$S$, $T$, $U$,
and the $Z\nu\nu$ coupling suppression parameter $\varepsilon$
on the electroweak observables.
The dependence of the observables on the oblique correction parameters was
obtained in Ref.~\cite{Peskin:1990zt} so we will not 
reproduce it here.
To obtain the dependence on $\varepsilon$, 
we note that the tree-level relation between
$\sin^2\theta_W$ and the input parameters $\alpha$, $G_F$, and $M_Z$ is
given by
\begin{equation}
\sin 2\theta = \left[ \frac{ 4\pi\alpha }{ \sqrt{2} G_F M_Z^2 }
               \right]^{1/2}\;.
\end{equation}
Thus, a shift in $G_F$, Eq.~(\ref{GFshift}), will shift 
$\sin^2\theta_W$ (be it
$\sin^2\theta_W^{\mathrm{(on-shell)}}$ or
$\sin^2\theta_\mathrm{eff}^\mathrm{lept}$) by
\begin{equation}
\delta s^2 
= \dfrac{s^2 c^2}{c^2-s^2}
  \left[ \dfrac{\delta\alpha}{\alpha}
        -\dfrac{\delta G_F}{G_F}
        -\dfrac{\delta M_Z^2}{M_Z^2}
  \right]
=  -\left(\dfrac{s^2 c^2}{c^2-s^2}\right)\varepsilon \;.
\end{equation}
Combining with the oblique corrections, we obtain
\begin{equation}
\begin{array}{l}
\delta\sin^2\theta_\mathrm{eff}^\mathrm{lept}
= \dfrac{\alpha S}{4(c^2-s^2)}
 -\dfrac{s^2 c^2}{c^2-s^2}\left( \alpha T + \varepsilon \right)\;, \\
\dfrac{\delta M_W}{M_W}
=  -\dfrac{\alpha S}{4(c^2-s^2)}
      +\dfrac{c^2}{2(c^2-s^2)}
       \left( \alpha T + \dfrac{s^2}{c^2}\varepsilon \right)
      +\dfrac{\alpha U}{8 s^2}\;.
\end{array}
\label{s2andMW}
\end{equation}
The $Z$ widths are proportional to the product
$\rho\,G_F M_Z^3$ \cite{Rosner:1993rj} and it is easy to see that
\begin{equation}
\dfrac{\delta(\rho\, G_F M_Z^3)}{\rho\, G_F M_Z^3}
= \alpha T + \varepsilon\;.
\label{Zwidthcorr}
\end{equation}
From Eqs.~(\ref{s2andMW}) and (\ref{Zwidthcorr})
we see that all $Z$-pole observables, 
except for the invisible width,
depend on $\varepsilon$ only through the combination
$(\alpha T + \varepsilon)$.  
Consequently, any shift in the $Z$-pole observables
due to a non-zero $\varepsilon$
could be compensated by a shift in the $T$ parameter
with the result that the observables remain unchanged.
Finally, the $Z$ invisible width must be corrected by
a factor of $(1-2\varepsilon)$ and the NuTeV
parameters $g_L^2$ and $g_R^2$ by a factor of $(1-\varepsilon)$.

Numerically, the observables are corrected as follows:
\begin{eqnarray}
\dfrac{  \Gamma_\mathrm{lept} }
      { [\Gamma_\mathrm{lept}]_\mathrm{SM} }
& = & 1 - 0.0021\,S + 0.0093\,T + 1.2\,\varepsilon\;, \cr
\dfrac{  \Gamma_\mathrm{inv}/\Gamma_\mathrm{lept} }
      { [\Gamma_\mathrm{inv}/\Gamma_\mathrm{lept}]_\mathrm{SM} }
& = & 1 + 0.0021\,S - 0.0015\,T - 2.2\,\varepsilon\;, \cr
\dfrac{  \sin^2\theta_{\mathrm{eff}}^{\mathrm{lept}} }
      { [\sin^2\theta_{\mathrm{eff}}^{\mathrm{lept}}]_\mathrm{SM} }
& = & 1 + 0.016\,S - 0.011\,T - 1.4\,\varepsilon\;, \cr
\dfrac{  g_L^2 }
      { [g_L^2]_\mathrm{SM} }
& = & 1 - 0.0090\,S + 0.022\,T - 0.17\,\varepsilon\;, \cr
\dfrac{  g_R^2 }
      { [g_R^2]_\mathrm{SM} }
& = & 1 + 0.031\,S - 0.0067\,T - 3.9\,\varepsilon\;, \cr
\dfrac{  M_W }
      { [M_W]_\mathrm{SM} }
& = & 1 - 0.0036\,S + 0.0056\,T \cr
&   & \qquad + 0.0042\,U + 0.22\,\varepsilon\;.
\label{STUeps}
\end{eqnarray}
Here, $[*]_\mathrm{SM}$ is the usual SM prediction of the 
observable $*$ using $G_\mu$ as input.
Note that the coefficient of $\varepsilon$ in $g_L^2$ is small.
This means that the dependence of $g_L^2$ on $\varepsilon$ due to the
suppression of the $Z\nu\nu$ and $W\ell\nu$ couplings is almost
canceled by the dependence of the SM prediction through $G_\mu$.
Ironically, although the original motivation for
introducing the parameter $\varepsilon$ was to fit $g_L^2$,
it turns out that $g_L^2$ is the least sensitive to $\varepsilon$
of all the observables considered.

\section{The Fits}

Using the data and formulae from the previous sections, we
perform several fits.  Initially, we exclude the $W$ mass and
perform a fit to the remaining five observables only since including
the $W$ mass serves only to determine the $U$ parameter and does
not affect the values of the other fit parameters.

\subsection{Oblique Corrections Only}

First, we perform a fit with only the oblique correction
parameters $S$ and $T$.
Using $M_\mathrm{top} = 174.3\,\mathrm{GeV}$,
$M_\mathrm{Higgs} = 115\,\mathrm{GeV}$ as the reference SM 
and fitting the expressions in Eq.~(\ref{STUeps}) to the three 
$Z$-pole observables
and the two NuTeV observables we obtain
\begin{eqnarray}
S & = & -0.09\pm 0.10\;, \cr
T & = & -0.13\pm 0.12\;,
\end{eqnarray}
with a correlation of $0.89$. The quality of the fit is 
unimpressive: $\chi^2=11.3$ for $5-2=3$ degrees of freedom. 
The preferred region on the $S$-$T$ plane is shown in Fig.~1.
As is evident from the figure, there is no region where the $1\sigma$ bands
for $\Gamma_\mathrm{lept}$, $\sin^2\theta_\mathrm{eff}^\mathrm{lept}$,
and $g_L^2$ overlap.

Fig.~1 also shows the preferred region if $S$ and $T$ are fit to the
$Z$-pole observables only.  Including the NuTeV data shifts
both central values of $S$ and $T$ to the negative side.
Though the $g_L^2$ band tries to pull $S$ and $T$ into the
fourth quadrant (which corresponds to the SM with a larger Higgs mass),
the narrow $\sin^2\theta_{\mathrm{eff}}^{\mathrm{lept}}$ 
band forces $S$ and $T$ to move along it
into the third quadrant.

In conclusion, oblique corrections alone cannot explain the NuTeV 
anomaly and/or the invisible width discrepancy.

\subsection{With Neutrino Mixing}

\begin{table}
\begin{tabular}{c|cccc}
\hline
    & $S$ & $T$ & $U$ & $\varepsilon$ \\
\hline
$S$\ & $\phantom{-}1.00$ &  &  &  \\
$T$\ & $\phantom{-}0.56$ & $\phantom{-}1.00$ & & \\  
$U$\ & $-0.20$ & $-0.73$ & $1.00$ & \\
$\varepsilon$\ & $\phantom{-}0.18$ & $-0.64$ & $0.58$ & $1.00$ \\
\hline
\end{tabular}
\caption{The correlations among the oblique correction parameters
and $\varepsilon$.  The correlations among $S$, $T$, and $\varepsilon$
are unaffected by whether $M_W$ and the $U$ parameter are included in the
fit.}
\label{STUepsCorr}
\end{table}

Next, we perform a fit with $S$, $T$, and $\varepsilon$.
The reference SM is $M_\mathrm{top} = 174.3\,\mathrm{GeV}$,
$M_\mathrm{Higgs} = 115\,\mathrm{GeV}$ as before.
The result is
\begin{eqnarray}
S & = & -0.03\pm 0.10\;, \cr
T & = & -0.44\pm 0.15\;, \cr
\varepsilon & = & 0.0030 \pm 0.0010\;.
\end{eqnarray}
The quality of the fit is improved dramatically to $\chi^2=1.17$ for
$5-3=2$ degrees of freedom.
The correlations among the parameters are shown in Table~\ref{STUepsCorr}.
The preferred regions in the $S$-$T$, $S$-$\varepsilon$, and $T$-$\varepsilon$
planes are shown in Figs. 2 through 4.
The central values have shifted somewhat from our preliminary 
report in Ref.~\cite{Takeuchi:2002nn}. This is partly
due to our use of updated NuTeV numbers from the third reference of
\cite{Zeller:2001hh}, and partly due to our use of
$\Delta\alpha_\mathrm{had}^{(5)}(M_Z) = 0.02761$ from
Ref.~\cite{Burkhardt:2001xp} instead of the ZFITTER default value of
$\Delta\alpha_\mathrm{had}^{(5)}(M_Z) = 0.02804$.

Thus, by including both oblique corrections and $\varepsilon$,
we obtain an excellent fit to both the $Z$-pole and NuTeV data.
But what kind of `new physics' would provide such values of $S$ and $T$?
While the limits on $S$ permit it to have either sign, $T$ is
constrained to be negative by $3\sigma$.  Few models of new physics 
are available which predict a negative $T$ \cite{Bertolini:1990ek}.

Recall that 
the effect of a SM Higgs heavier than our reference value (here chosen
to be $115\,\mathrm{GeV}$) is manifested as shifts in the oblique correction
parameters.
The approximate expressions for these shifts are \cite{Takeuchi:1991kf}
\begin{eqnarray}
S_\mathrm{Higgs} & \approx & \frac{1}{6\pi}
     \ln\left(M_\mathrm{Higgs}/M_\mathrm{Higgs}^\mathrm{ref}\right)\;, \cr
T_\mathrm{Higgs} & \approx & -\frac{3}{8\pi c^2}
     \ln\left(M_\mathrm{Higgs}/M_\mathrm{Higgs}^\mathrm{ref}\right)\;, \cr
U_\mathrm{Higgs} & \approx & 0\;.
\label{HiggsSTU}
\end{eqnarray}
Notice that $S_\mathrm{Higgs}$ is positive while 
$T_\mathrm{Higgs}$ is negative for
$M_\mathrm{Higgs} > M_\mathrm{Higgs}^\mathrm{ref}$.
Thus increasing the Higgs mass will have the desired effect of
providing a negative $T$.  Can $T$ be made negative enough
without making $S$ too large?
The answer is provided in Fig.~5, in which we show a blow up of the 
central region of Fig.~2.
The SM points fall comfortably within the 90\% confidence contour
when the Higgs mass is even moderately large.

\subsection{Neutrino Mixing Only}

To check that the SM itself is indeed compatible with the
data (excluding the $W$ mass) we perform a fit with only $\varepsilon$ as the
fit parameter, and plot the dependence of $\chi^2$ on 
the Higgs mass used to define the reference SM.
The result is shown is Fig.~6. We also show the $M_\mathrm{Higgs}$
dependence of the $1\sigma$ limits on $\varepsilon$ in Fig.~7.  
The graphs demonstrate that the data are well fit by 
a SM with a heavy Higgs and a nonzero value of $\varepsilon$.

\subsection{The $W$ Mass}

We have deliberately excluded the $W$ mass from our analysis since it
cannot be fit with $\varepsilon$ alone, or by $S$, $T$, and $\varepsilon$.
Eq.~(\ref{s2andMW}) suggests that the single parameter fit with 
$\varepsilon$, which yields a positive value of $\varepsilon$, 
may increase the $W$ mass prediction towards the
experimental value \cite{Babu:1991ny}.  However, 
the coefficient of $\varepsilon$, 
Eq.~(\ref{STUeps}), is too small for it to significantly 
mitigate the discrepancy.
A three parameter fit with $S$, $T$, and $\varepsilon$ also fails to
close the gap.  
In that case, note that the combination of $T$ and $\varepsilon$
upon which $M_W$ depends is
\begin{equation}
\left(\alpha T + \frac{s^2}{c^2}\varepsilon\right)
= \left( \alpha T + \varepsilon \right) - 
  \left( \frac{c^2-s^2}{c^2} \right) \varepsilon\;.
\end{equation}
Since $(\alpha T + \varepsilon)$ is pinned by the $Z$-pole 
observables, a non-zero $\varepsilon$ would actually lead to a 
\textit{decrease} in the $W$ mass prediction.
Therefore, the $U$ parameter is necessary to fit $M_W$.
The result of the four parameter fit to the reference SM of
$M_\mathrm{top} = 174.3\,\mathrm{GeV}$, 
$M_\mathrm{Higgs} = 115\,\mathrm{GeV}$ is
\begin{equation}
U = 0.62 \pm 0.16\;,
\end{equation}
and the correlations with the other parameters are shown in 
Table~\ref{STUepsCorr}.  The dependence of these limits on
the reference Higgs mass $M_\mathrm{Higgs}$ is weak, as implied by
Eq.~(\ref{HiggsSTU}) and shown in Fig.~8.

\section{Conclusions and Discussion}

We conclude that neutrino mixing together with oblique corrections 
can reconcile the $Z$-pole and NuTeV data.  The simplest model
which provides the necessary values of $S$ and $T$ is the SM itself
with a large Higgs boson mass.

We emphasize that there is nothing exotic about this solution.
Neutrinos are known to mix from Super-Kamiokande \cite{SK}, 
SNO \cite{SNO}, K2K \cite{K2K}, and other neutrino experiments.
Neutrino mixing may have the added bonus of contributing 
negatively to both $S$ and $T$ \cite{Bertolini:1990ek}.

The possibility that the Higgs boson is heavy 
has also been under scrutiny \cite{Peskin:2001rw,Chanowitz:2001bv}
since it has not been found in the $\sim 80$~GeV range 
preferred by the SM global fit (Ref.~\cite{LEP/SLC:2002}, page 154).
If the Higgs boson is indeed heavy as suggested by Fig.~6,
it may favor a possible dynamical mechanism for electroweak symmetry 
breaking while providing a challenge for supersymmetry \cite{SUSY}.

The limits on the suppression parameter,
\begin{equation}
\varepsilon = 0.0030 \pm 0.0010\;,
\end{equation}
imply
\begin{equation}
\theta = 0.055 \pm 0.010\;,
\end{equation}
as the mixing angle if the suppression is due to mixing with a single 
heavy sterile state.
The heavy state into which the neutrino mixes
must be at least as massive as the $Z$ so that the $Z$ does not
decay into it.   A naive seesaw model \cite{seesaw} does not permit 
such a large mixing angle since the angle is related to the 
ratio of masses.
However, it can be shown that the required pattern of mixings and masses can 
be arranged when there exist inter-generational mixings \cite{Chang:1994hz}.
We will present explicit examples in Ref.~\cite{LOTW2}.

The $W$ mass needs the $U$ parameter, so we cannot do without
other new physics entirely.   The Higgs mass dependence of the
oblique correction parameters are shown in Fig.~8.
What kind of new physics will be compatible with this?
It has to predict a small $T$ while predicting a large $U$.
One possibility is that the $U$ parameter can be enhanced
by the formation of bound states at new particle thresholds.
If one expresses $T$ and $U$ as dispersion integrals over
spectral functions, one finds
\begin{eqnarray}
T & \propto & \int_{s_\mathrm{thres}}^\infty \frac{ds}{s}
              \left[ \mathrm{Im}\Pi_\pm(s) - \mathrm{Im}\Pi_0(s)
              \right]\;, \cr
U & \propto & \int_{s_\mathrm{thres}}^\infty \frac{ds}{s^2}
              \left[ \mathrm{Im}\Pi_\pm(s) - \mathrm{Im}\Pi_0(s)
              \right]\;,
\end{eqnarray}
where we have used the notation of Ref.~\cite{Takeuchi:1994ym}.
Due to the extra negative power of $s$ in the integrand of $U$,
it is more sensitive to the enhancement of the threshold than $T$.
Indeed, it has been shown in Ref.~\cite{Takeuchi:1994ym} that
threshold effects do not enhance the $T$ parameter.
This could, again, mean that a theory that exhibits dynamical 
electroweak symmetry breaking is favored.
This possibility will be investigated in a subsequent paper.

\section*{Acknowledgments}

We would like to thank Lay Nam Chang, Michael Chanowitz, 
Michio Hashimoto, Randy Johnson, Alex Kagan, Kevin McFarland, 
Mihoko Nojiri, Jogesh Pati, and Mike Shaevitz
for helpful discussions and communications.
This research was supported in part by the U.S. Department of Energy, 
grants DE--FG05--92ER40709, Task A (T.T. and N.O.),
and DE-FG02-84ER40153 (L.C.R.W.).


\widetext
\newpage

\begin{figure}[p]
\begin{center}
\scalebox{0.75}{\includegraphics{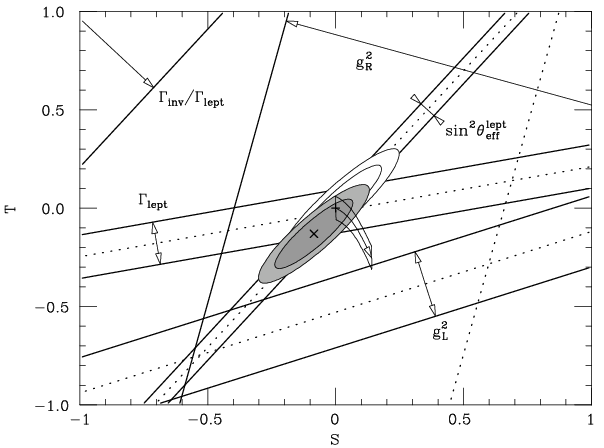}}
\caption{The fit to the data with only $S$ and $T$. 
The bands associated with each observable show the $1\sigma$ limits.
The shaded ellipses show the 68\% and 90\% confidence contours. 
The unshaded ellipses partially hidden behind the shaded ones show the 
contours when only the LEP/SLC data is used.  
The origin is the reference SM with $M_\mathrm{top}=174.3\,\mathrm{GeV}$ and
$M_\mathrm{Higgs}=115\,\mathrm{GeV}$.  The curved arrow attached to
the origin indicates the path along which the SM point will move when the
Higgs mass is increased from 115~GeV to 1~TeV.  The curved lines above
and below it indicate the uncertainty in the SM point's position from
that of the top mass: $M_\mathrm{top}=174.3\pm 5.1\,\mathrm{GeV}$ 
\protect{\cite{CDF/D0:2000}}.
}
\label{fig1}
\end{center}
\end{figure}

\begin{figure}[p]
\begin{center}
\scalebox{0.75}{\includegraphics{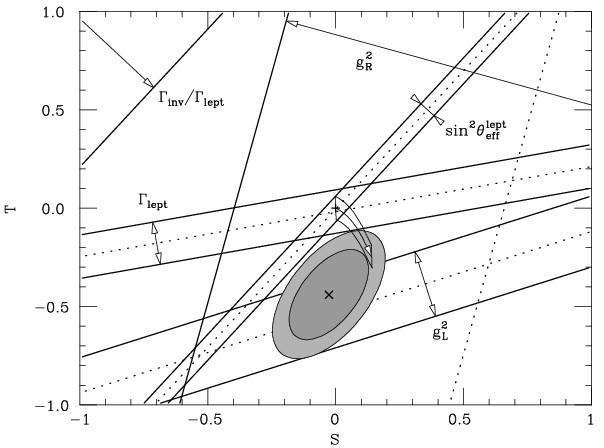}}
\caption{The 68\% and 90\% confidence contours on the $S$-$T$ plane
when the data is fit with $S$, $T$, and $\varepsilon$.
The SM points are shown as in Fig.~1.}
\label{fig2}
\end{center}
\end{figure}

\begin{figure}[p]
\begin{center}
\scalebox{0.75}{\includegraphics{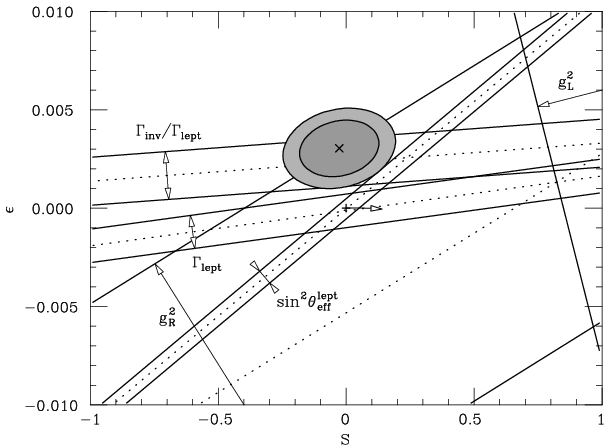}}
\caption{The 68\% and 90\% confidence contours on the $S$-$\varepsilon$ plane.
The arrow attached to the origin indicates the path along which the
SM point will move when $M_\mathrm{Higgs}$ is increased from 115~GeV to 1~TeV.
The dependence of the SM point on $M_\mathrm{top}$ is not shown.}
\label{fig3}
\end{center}
\end{figure}

\begin{figure}[p]
\begin{center}
\scalebox{0.75}{\includegraphics{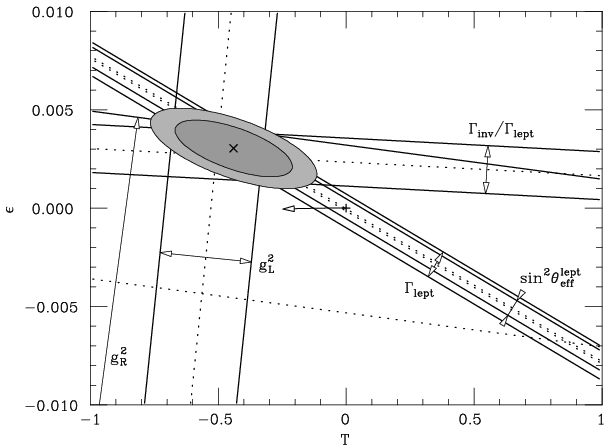}}
\caption{The 68\% and 90\% confidence contours on the $T$-$\varepsilon$ plane.
The arrow attached to the origin indicates the path along which the
SM point will move when $M_\mathrm{Higgs}$ is increased from 115~GeV to 1~TeV.
The dependence of the SM point on $M_\mathrm{top}$ is not shown.}
\label{fig4}
\end{center}
\end{figure}

\begin{figure}[p]
\begin{center}
\scalebox{0.75}{\includegraphics{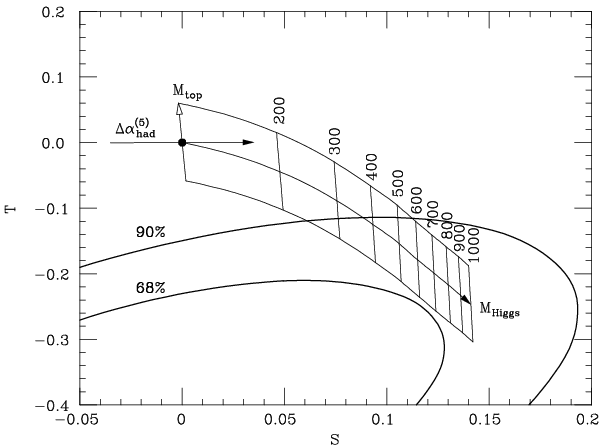}}
\caption{Blow up of the central region of Fig.~2.
The $M_\mathrm{Higg}$ and $M_\mathrm{top}$ dependence of the SM point
is shown as in Fig~2, with the Higgs mass indicated in units of GeV.
The horizontal arrow going through the origin indicates the uncertainty
in the position of the SM point due to the uncertainty in 
$\Delta\alpha_\mathrm{had}^{(5)}(M_Z)$, with the arrow pointing in the
direction of increasing $\Delta\alpha_\mathrm{had}^{(5)}(M_Z)$}
\label{fig5}
\end{center}
\end{figure}

\begin{figure}[p]
\begin{center}
\scalebox{0.75}{\includegraphics{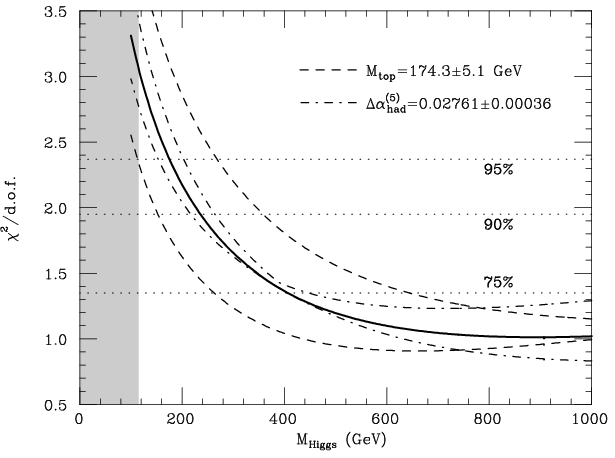}}
\caption{The dependence of $\chi^2$/d.o.f. on $M_\mathrm{Higgs}$ 
when only $\varepsilon$ is used to fit the $Z$-pole and NuTeV data.
The $W$ mass is not included in the fit.
The Higgs mass in the shaded region is excluded by direct searches.}
\label{fig6}
\end{center}
\end{figure}

\begin{figure}[p]
\begin{center}
\scalebox{0.75}{\includegraphics{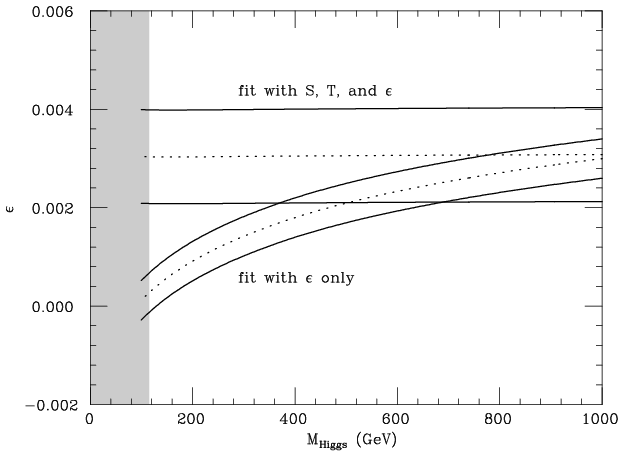}}
\caption{The dependence of the $1\sigma$ limits of $\varepsilon$
on $M_\mathrm{Higgs}$.
The Higgs mass in the shaded region is excluded by direct searches.}
\label{fig7}
\end{center}

\end{figure}
\begin{figure}[p]
\begin{center}
\scalebox{0.75}{\includegraphics{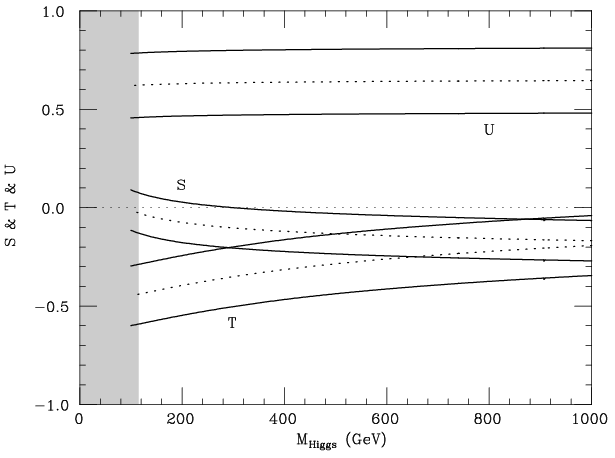}}
\caption{The dependence of the $1\sigma$ limits of $S$, $T$, and $U$ 
on $M_\mathrm{Higgs}$.
The Higgs mass in the shaded region is excluded by direct searches.}
\label{fig8}
\end{center}
\end{figure}

\end{document}